\documentclass[twocolumn, pre, showpacs, english, preprintnumbers, amsmath, amssymb, superscriptaddress, aps,longbibliography]{revtex4-2}
\usepackage[subpreambles=true]{standalone}

\usepackage[utf8]{inputenc}
\usepackage{graphicx}
\usepackage{grffile}
\usepackage[usenames,dvipsnames]{xcolor}
\usepackage{amsmath}
\usepackage[normalem]{ulem}
\usepackage[resetlabels, labeled]{multibib}
\usepackage{appendix}
\newcites{S}{References Supplementary Materials}
\definecolor{orange}{rgb}{1,0.5,0}
\definecolor{goodgreen}{rgb}{0.1,0.5,0}
\definecolor{goodred}{rgb}{0.7,0,0}
\usepackage{lineno}
\usepackage{tikz}
\usepackage{tocvsec2}
\usepackage{scrextend}
\usepackage{lineno}
\makeatletter

\usepackage[colorlinks,urlcolor=goodgreen,citecolor=blue,linkcolor=goodred]{hyperref}

\usepackage{float}
\graphicspath{ {images/} }

\let\oldepsilon\epsilon \let\epsilon\varepsilon \let\varepsilon\oldepsilon

\makeatother

\usepackage{babel}

\begin{document}

%\title{Superconductor/ferromagnetic insulator bilayer with arbitrary domain magnetization}
\title{Quasiparticle density of states and triplet correlations in superconductor/ferromagnetic-insulator structures across a sharp domain wall  }
\newcommand{\orcid}[1]{\href{https://orcid.org/#1}{\includegraphics[width=8pt]{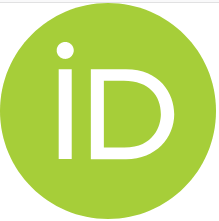}}}

\author{Alberto Hijano\orcid{0000-0002-3018-4395}}
\email{alberto.hijano@ehu.eus}
\affiliation{Centro de F\'isica de Materiales (CFM-MPC) Centro Mixto CSIC-UPV/EHU, E-20018 Donostia-San Sebasti\'an,  Spain}
\affiliation{Department of Condensed Matter Physics, University of the Basque Country UPV/EHU, 48080 Bilbao, Spain}

%\author{Stefan Ili\'{c}}
%\email{stefan.ilic@csic.es}
%\affiliation{Centro de F\'isica de Materiales (CFM-MPC) Centro Mixto CSIC-UPV/EHU, E-20018 Donostia-San Sebasti\'an,  Spain}

\author{Vitaly N. Golovach\orcid{0000-0001-7457-171X}}
\email{vitaly.golovach@ehu.eus}
\affiliation{Centro de F\'isica de Materiales (CFM-MPC) Centro Mixto CSIC-UPV/EHU, E-20018 Donostia-San Sebasti\'an,  Spain}
\affiliation{Donostia International Physics Center (DIPC), 20018 Donostia--San Sebasti\'an, Spain}
\affiliation{IKERBASQUE, Basque Foundation for Science, 48011 Bilbao, Spain}
\author{F. Sebasti\'{a}n Bergeret\orcid{0000-0001-6007-4878}}
\email{fs.bergeret@csic.es}
\affiliation{Centro de F\'isica de Materiales (CFM-MPC) Centro Mixto CSIC-UPV/EHU, E-20018 Donostia-San Sebasti\'an,  Spain}
\affiliation{Donostia International Physics Center (DIPC), 20018 Donostia--San Sebasti\'an, Spain}

\begin{abstract}
A ferromagnetic insulator (FI) in contact with a superconductor (S) is known to induce a spin splitting of the BCS density of states at the FI/S interface. This spin splitting causes the Cooper pairs to reduce their singlet-state correlations and acquire odd-in-frequency triplet correlations. We consider a diffusive FI/S bilayer with a sharp magnetic domain wall in the FI and study the local quasiparticle density of states and triplet superconducting correlations. 
In the case of collinear alignment of the domains, we obtain analytical results by solving the Usadel equation.   For a small enough exchange field or weak superconductivity, we also find an analytical expressions for arbitrary magnetic textures, which reveals how the triplet component vector depends on the local magnetization of the FI. 
For an arbitrary angle between the magnetizations and the strength of the exchange field, we numerically solve the problem of a sharp domain wall. We finally propose two different setups based on  FI/S/F stacks, where F is a ferromagnetic layer, to filter out singlet pairs and detect the presence of triplet correlations via tunneling differential conductance measurements. 
%The spectral properties of a superconductor (S)  attached to a ferromagnetic insulator (FI) are considerably affected by the interfacial exchange field. For example, if S is a film with a thickness smaller than the superconducting coherence length, the exchange field is manifested in a homogeneous  spin-splitting of the quasiparticle spectrum. Another consequence of the exchange field,  not studied as much in this context, is the appearance of superconducting triplet correlations. In this work, we study the spectral properties of a diffusive FI/S bilayer in the presence of a sharp magnetic domain wall.  We present an analytical solution of the Usadel equation and determine the electronic spectral properties of a FI/S bilayer with  two magnetic domains.  If the domains have antiparallel magnetization we find an analytical solution for the quasiparticle density of states.  In the case of non-collinear domains we solve the equations  numerically and calculate the spatial dependence of the different components of the condensate.   Finally, we propose a simple setup based on FI/S bilayer  that allows the detection of triplet components through tunneling spectroscopy. 
\end{abstract}

\maketitle

\section{Introduction}

The exchange coupling at the interface between a ferromagnetic insulator (FI)  and a thin superconducting layer (S) can lead to a spin-splitting of the density of states (DoS) in the S layer, as observed in numerous experiments~\cite{moodera1988electron,Hao:1990,meservey1994spin,Strambini-2017,Hijano:2021}. Recently there has been a renewed interest in these systems due to various proposed applications. These applications include  spin valves~\cite{miao2014spin,DeSimoni:2018}, spin batteries~\cite{Jeon:2020,Ojajarvi:2021}, magnetometers~\cite{Alidoust:2013,strambini2015mesoscopic}, thermometers~\cite{Giazotto:2006,Giazotto:2015}, caloritronic devices~\cite{giazotto2013phase,giazotto2015very,giazotto2020very}, thermoelectric elements~\cite{machon2013nonlocal,ozaeta2014predicted}, and radiation detectors~\cite{Heikkila:2018,Geng:2020}. FI/S structures have also been explored in the context of  Majorana fermions in semiconducting wires~\cite{Oreg:2010,lutchyn2010majorana,Dai:2021}.

Most of these applications require a robust superconducting gap with a sizable spin-splitting. This can be achieved,   for example,  in  EuS/Al systems, ~\cite{Hao:1990,Moodera:2007,Strambini-2017,DeSimoni:2018,Hijano:2021}, where the interfacial exchange interaction leads to a sharp spin-splitting in S layers with thicknesses smaller than the coherent length.  On the theoretical side,   the effect of the interfacial exchange field and the induced spin-splitting on the superconducting state has been studied in numerous works~\cite{Tokuyasu:1988,Virtanen:2020,rouco2019charge,Hijano:2021}.  Most of these works assume a homogeneous spin-splitting field. This assumption is justified,  even in a multidomain situation,  if the characteristic domain size of EuS is much longer than the superconducting coherence length $\xi_0$.

There are, however, situations in which the domain size may be of the order of the superconducting coherence length. The effect of  domain walls in magnetic and insulating ferromagnets on adjacent superconductors has been studied  theoretically~\cite{Aladyshkin:2003,Houzet:2006,Bobkova:2019,Rabinovich:2019,Aikebaier-2019} and experimentally~\cite{Yang:2004,Strambini-2017}, while Ref.~\cite{Linder-2014} studied the influence of domain-wall dynamics on superconductivity. In particular, Ref.~\cite{Strambini-2017} provided experimental evidence that EuS consists of multiple domains with a size of the order of the coherence length of the  Al layer attached to it.  The authors of that work contrast spectroscopic measurements with a theoretical model that assumed alternating up/down domains of different sizes. 
In the present work, we generalize this approach and study a FI/S structure with two non-collinear magnetic domains.

Despite the amount of experimental work on FI/S systems, almost all of it focuses on studying its quasiparticle spectrum. There is, however, an interesting aspect that is not often mentioned in these works. The mere existence of an interfacial exchange field leads to conversion of singlet superconducting correlations to triplet ones~\cite{Bergeret:2005,cottet2011inducing,bergeret2018colloquium,heikkila2019thermal}.   The induced triplet component has a total zero spin projection if the FI  consists of a single domain with homogeneous magnetization.    However,  in FI/S systems with non-collinear magnetization, triplet components with different spin-projections may coexist with the singlet one.

 In this work, we study the equilibrium properties of a  FI/S bilayer with a sharp domain wall separating two magnetic domains. We present an analytical solution for the Usadel equation for a FI/S bilayer consisting of two semi-infinite magnetic domains with collinear magnetization and non-collinear magnetization in the weak exchange field limit; and use numerical methods to solve the non-collinear case with arbitrary exchange field strength. Additionally, we study the spatial evolution of the triplet correlations near a domain wall and propose a method to detect them using tunneling spectroscopy of an additional ferromagnetic layer.  The work is organized as follows:  In Sec.~\ref{model and formalism} we present the main equations describing a diffusive superconductor attached to a FI layer with multiple domains and a general Lagrangian from which one can derive the Usadel equation.   We identify conserved quantities within each domain. In Sec.~\ref{collinear magnetization} we use these integrals of motion to derive an analytical expression for the DoS of a  FI/S system with two collinear domains of arbitrary magnitude.  In Sec.~\ref{arbitrary magnetization} we generalized these results to the case of non-colinear magnetization.  Finally, in Sec.~\ref{entangled triplet pairs} we study the properties and spatial evolution of the triplet correlations,  and suggest a way to detect them.   We summarize the results in Sec.~\ref{conclusion}.

%In Appendix~\ref{tunneling differential conductance} we show that the long-range triplet correlations lead to a positive correction of the DoS at zero-energy in a ferromagnet attached to the S layer.

\section{The Model}\label{model and formalism}

\begin{figure}[!t]
    \centering
    \includegraphics[width=0.99\columnwidth]{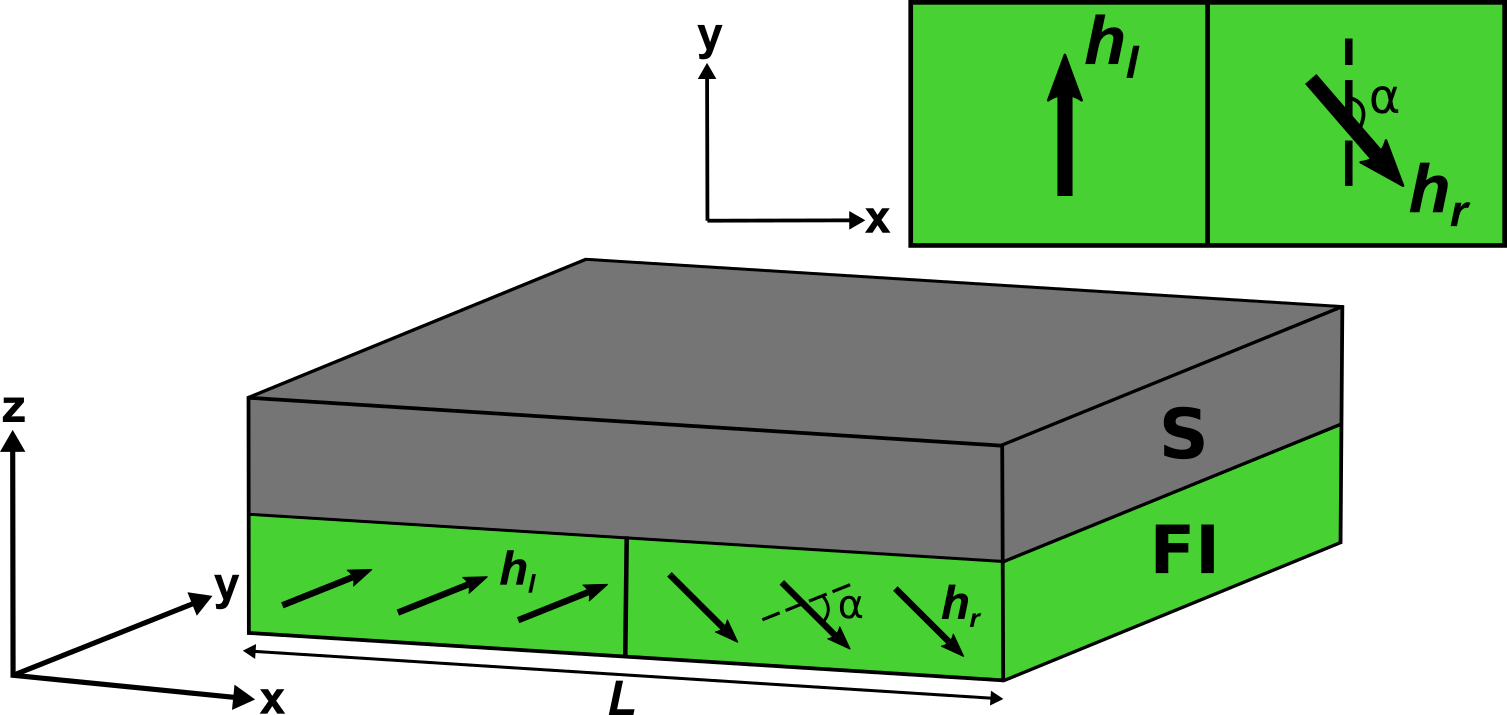}
    \caption{Schematic view of the S/FI structure under consideration. The  ferromagnetic insulator has two domains with arbitrary in-plane magnetization direction. The inset shows the top view of the FI. The magnetizations of the two domains lie on the $xy$ plane, and form and angle $\alpha$.}\label{S-FI schematic}
\end{figure}

We consider a FI/S bilayer structure, see Fig.~\ref{S-FI schematic}. A diffusive superconducting film is placed on top of a FI film.
A typical example is Eu/Al studied in several papers~\cite{Hao:1990,Moodera:2007,Strambini-2017,DeSimoni:2018,Hijano:2021}. In these systems, the EuS film is polycrystalline and magnetic domains with sharp boundaries are very common, in particular before the first magnetization of the EuS~\cite{Strambini-2017}.
 
%From the experimental data of Ref. \cite{Strambini-2017} the 
%\AHedit{Before magnetization, the multi domain structure of a polycrystalline FI layer leads to an inhomogeneous magnetic texture~\cite{Strambini-2017,Hijano:2021}. Domain walls are not rare, and contribute sizeably, for example, to the tunneling spectroscopy.} First, we assume that the FI film has an arbitrary  magnetic texture. The magnetic proximity effect leads to an exchange field in the S layer, and as a consequence there is a spin-splitting of the DoS.

%In the nonlinear $\sigma$-model, the action of a disordered superconductor subject to an exchange field is given by~\cite{Kamenev-2011}
%\begin{equation}\label{action}
%    S[Q]=\frac{\pi N_0}{4} \mathrm{Tr}\{D/2(\partial_x Q)^2+2(i\epsilon\tau_3-i\boldsymbol{h}\cdot\boldsymbol{\sigma}\tau_3-\check{\Delta})Q\}\; ,
%\end{equation}
%where the matrix field $Q$ satisfies the normalization condition $Q^2=1$. Here, Tr includes the matrix trace as well as integration over time and position, $\epsilon$ is represented in the time representation $\epsilon=i\partial_t \delta(t-t')$, $D$ is the diffusion constant, $\boldsymbol{h}$ is the exchange field, $\check{\Delta}=\Delta\tau_1$ is the order parameter and $N_0$ is the DoS at the Fermi-level. 

%%%%%%%START FIRST WITH GENERAL EXCHANGE FIELD%%%%

To describe the system we use  the quasiclassical Green’s function (GF) formalism extended to treat spin-dependent fields ~\cite{Bergeret:2005}.
In this case  the GF $\check{g}$ is a $4 \times 4$ matrix in Nambu-spin space.  In the diffusive limit, it  does not depend on momentum and is determined   by the Usadel equation~\cite{Usadel}. The interfacial exchange field is introduced as an effective boundary condition at the FI/S interface~\cite{bergeret2012electronic,eschrig2015general,heikkila2019thermal}. Assuming that the thickness of the S layer is smaller than the coherence length one can integrate the Usadel equation over the thickness to reduce the dimension of the problem. The resulting  Usadel equation for the retarded GF reads
\begin{equation}\label{usadel_equation}
    D \nabla\cdot(\check{g}\nabla\check{g})+[i\epsilon\tau_3-i\boldsymbol{h}\cdot\boldsymbol{\sigma}\tau_3-\check{\Delta},\check{g}]=0\; ,
\end{equation}
where $\nabla=(\partial_x,\partial_y)$,   $D$ is the diffusion constant, $\epsilon$ is the energy, $\boldsymbol{h}$ is the effective exchange field stemming from the interface and $\check{\Delta}=\Delta\tau_1$ is the order parameter. The exchange field is only finite at the FI/S interface, and we approximate it as  $|\boldsymbol{h}_{\mathrm{int}}|=h_{\mathrm{int}} a\delta(z)$, where $h_{\mathrm{int}}(x,y)$ is the exchange field at the interface, and $a$ is the thickness of an effective layer over which the exchange interaction is finite~\cite{heikkila2019thermal,Zhang:2019}. After integration over the $z$ direction the effective exchange  field is given by $h=h_{\mathrm{int}}a/d$~\cite{Hijano:2021}. It is worth noting that the critical temperature of the S layer decreases with decreasing thickness, such that at low temperatures superconductivity can be fully suppressed  when  $h > \Delta/\sqrt{2}$~\cite{Chandrasekhar:1962,Clogston:1962}. In this work we consider values of the exchange field  which are weak enough such that superconducting ordering and the exchange field coexist. The matrices $\sigma_i$ ($\tau_i$), $i=1,2,3$ in Eq.~\ref{usadel_equation} are the Pauli matrices in the spin (Nambu) space. The general structure of $\check g$ is:
\begin{equation}
    \label{eq:gen_str_g}
    \check g=\hat g\tau_3+\hat f\tau_1\; ,
\end{equation}
where $\hat g$ and $\hat f$ are the normal and anomalous GF in spin-space.

The GF  satisfies the normalization condition $\check{g}^2=1$, and can be parametrized with the help of the  generalized $\theta$-parametrization~\cite{Ivanov-2006}
\begin{equation}\label{parametrization}
	\check{g}=(\cos{\theta}V_0 -\sin{\theta}\boldsymbol{V}\cdot\boldsymbol{\sigma})\tau_3+(\sin{\theta}V_0 + \cos{\theta}\boldsymbol{V}\cdot\boldsymbol{\sigma})\tau_1\; ,
\end{equation}
which is described by  two scalars $\theta$ and $V_0$  and the  vector $\boldsymbol{V}$. $V_0$ and $\boldsymbol{V}$ satisfy the  condition
\begin{equation}\label{normalization}
	V_0^2+\boldsymbol{V}^2=1\; .
\end{equation}
$V_0$ and $\boldsymbol{V}$ describe the singlet and triplet correlations respectively. If $\boldsymbol{h}$ is homogeneous then $\boldsymbol{V}$ is parallel to it, but in general, as we show on Sec.~\ref{entangled triplet pairs},  $\boldsymbol{V}$ is not  parallel to the local exchange field.

In the above parametrization,  the Usadel equation reduces to following set of equations
\begin{subequations}\label{Usadel equation 2}
\begin{align}
	& D\nabla^2 \theta+2i\epsilon\sin{\theta}V_0-2i\cos{\theta}\boldsymbol{h}\cdot\boldsymbol{V}+2\Delta\cos{\theta}V_0=0\\
	\begin{split}
	    &D (V_0 \nabla^2\boldsymbol{V}-\boldsymbol{V}\nabla^2 V_0) +2i\epsilon\cos{\theta}\boldsymbol{V}-2i\sin{\theta}\boldsymbol{h} V_0\\
	    & \hspace{130pt} -2\Delta\sin{\theta}\boldsymbol{V}=0\; .
	\end{split}
\end{align}
\end{subequations}
%The components of the Usadel equation~\eqref{Usadel equation 2} can also be obtained from the Lagrangian corresconding to the action~\eqref{action}. For example, parametrizing $Q$ in the generalized $\theta$-parametrization~\eqref{parametrization}, the Lagrangian is given by:
It is useful for finding analytical solutions to  write a Lagrangian which leads to  Equations~\eqref{Usadel equation 2} as  the Euler-Lagrange equations:
\begin{multline}\label{Lagrangian theta}
	\mathcal{L}=\frac{D}{2}\sum_{\mu}(\nabla V_\mu)^2+\frac{D}{2}(\nabla\theta)^2+2i\epsilon\cos{\theta}V_0\\
	+2i\sin{\theta}\boldsymbol{h}\cdot\boldsymbol{V}-2\Delta\sin{\theta}V_0\; ,
\end{multline}
with $\mu=0, 1, 2, 3$. This  Lagrangian coincides with the form of the non-linear $\sigma$-model from which the Usadel equation can also be derived~\cite{Kamenev-2011,Altland-2010-CondensedMatter}.

\begin{figure*}[!t]
    \centering
    \includegraphics[width=0.9\textwidth]{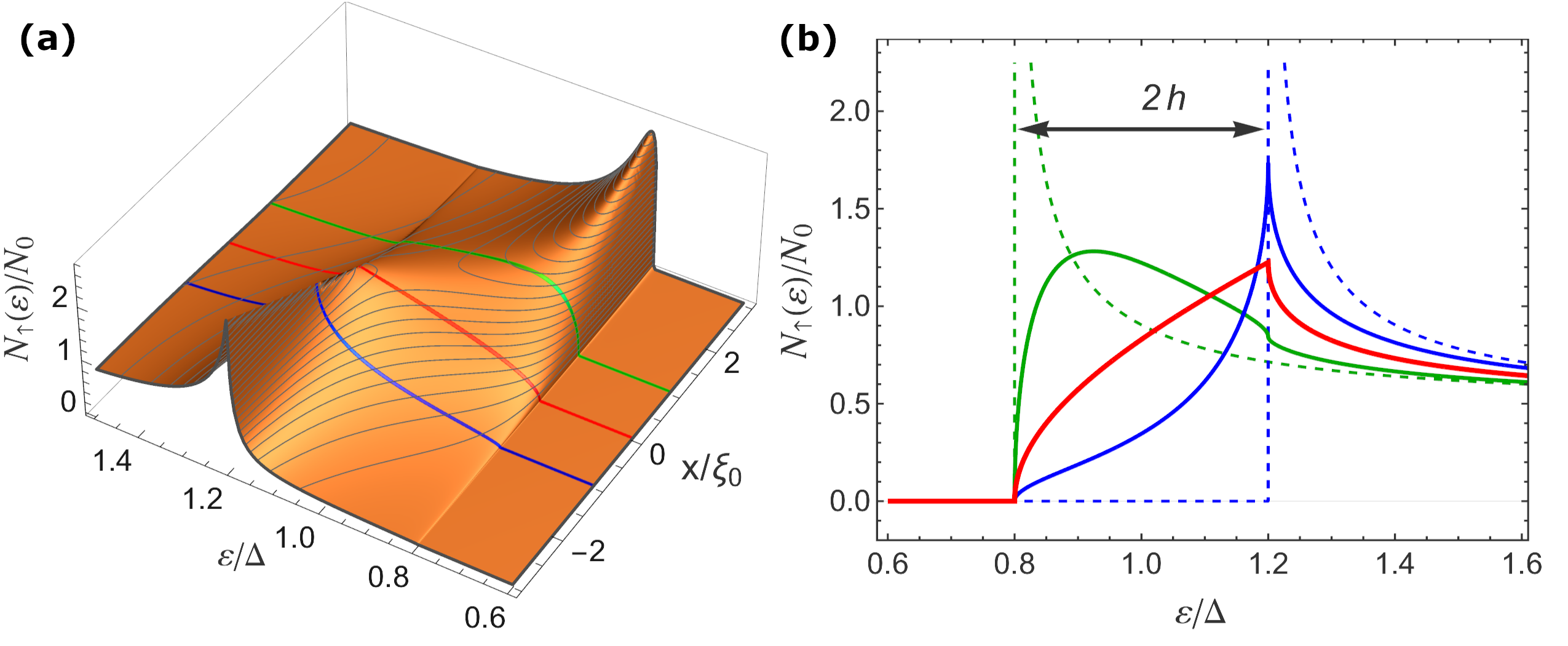}
    \caption{Local DoS (for spin-up) of the superconductor film for domains with opposite magnetization strength and effective exchange field $h=0.2\Delta$. The line traces of the right panel are taken at $x=-\xi_0$ (blue), $x=0$ (red) and $x=\xi_0$ (green). The dashed lines show the BCS spin-splitting of the DoS deep inside of the domains $x \rightarrow -\infty$ (blue) and $x \rightarrow \infty$ (green).}\label{DOS plot}
\end{figure*}

The above equations are valid for arbitrary magnetic textures.  On the following, we focus on the situation of  two semi-infinite magnetic domains with constant magnetization. The domains are separated by a sharp domain wall at $x=0$ with a length much smaller than the superconducting coherence length.  We assume that one of the  domains ($x<0$) is polarized along the $y$ axis, whereas the magnetization of the other domain  ($x>0$) forms an angle $\alpha$ with the $y$ axis, see Fig.~\ref{S-FI schematic}. At distances much larger than the coherence length  the GF takes its bulk form .  The system has translational symmetry along the $y$ and $z$ directions, so the parameters only depend on the $x$ coordinate.

%On the contrary, near the domain wall, the effect of one domain penetrates into the other domain over the superconducting coherence length. This can be verified by probing the spin-resolved DoS over the length of the superconductor. The exchange field is only finite at the FI/S interface, and we approximate it as  $|\boldsymbol{h}|=h_{\mathrm{int}} a\delta(z)$, where $h_{\mathrm{int}}(x,y)$ is the exchange field at the interface, and $a$ is the thickness of an effective layer over which the exchange interaction is finite~\cite{heikkila2019thermal,Zhang:2019}. Assuming that the thickness $d$ of the S layer is much smaller than the coherence length, we may integrate the Usadel equation over the $z$ direction and replace $\boldsymbol{h}$ with an effective field $h=h_{\mathrm{int}}a/d$~\cite{Hijano:2021} which is homogeneous along the $z$ direction. 

On each domain the Lagrangian~\eqref{Lagrangian theta} does not depend explicitly on the position $\boldsymbol{r}$, so the corresponding Hamiltonian is an integral of motion in each domain. The conserved quantity is namely  given by
\begin{multline}
	\mathcal{E}=\frac{D}{2}\sum_{\mu}(\nabla V_\mu)^2+\frac{D}{2}(\nabla\theta)^2-2i\epsilon\cos{\theta}V_0\\
	-2i\sin{\theta}\boldsymbol{h}\cdot\boldsymbol{V}+2\Delta\sin{\theta}V_0\; .
\end{multline}
The values  of $\mathcal{E}$ far away from the domain wall, where the GF is given by the bulk solution and therefore  constant in space can be easily obtained: 
\begin{equation}
	\mathcal{E}=-2i\epsilon\cos{\bar{\theta}}\bar{V}_0-2i\sin{\bar{\theta}}\boldsymbol{h}\cdot\bar{\boldsymbol{V}}+2\Delta\sin{\bar{\theta}}\bar{V}_0\; ,
\end{equation}
where the bulk values $\bar{\theta}$ and $\bar{\boldsymbol{V}}$ of the GF are given by the inverse relations of Eq.~\eqref{eq:gen_str_g}
\begin{subequations}\label{bulktheta}
    \begin{align}
        \tan{\theta}&=\frac{f_0}{g_0}\\
        \boldsymbol{V}&=-\frac{\boldsymbol{g}}{\sin{\theta}}\\
        V_0&=\frac{g_0}{\cos{\theta}}\; ,
    \end{align}
\end{subequations}
and the bulk GF is given by
\begin{subequations}\label{bulk g,f}
\begin{align}
	\hat{\overline{g}}&=\frac{-i(\epsilon-\boldsymbol{h}\cdot\boldsymbol{\sigma})}{\sqrt{\Delta^2-(\epsilon-\boldsymbol{h}\cdot\boldsymbol{\sigma})^2}}\\
	\hat{\overline{f}}&=\frac{\Delta}{\sqrt{\Delta^2-(\epsilon-\boldsymbol{h}\cdot\boldsymbol{\sigma})^2}}\; .
\end{align}
\end{subequations}
with $\hat{g}=g_0+\boldsymbol{g}\cdot\boldsymbol{\sigma}$, $\hat{f}=f_0+\boldsymbol{f}\cdot\boldsymbol{\sigma}$.
In the following section we use this expressions to define integrals of motion that allow for an analytical solution when the magnetic domains are collinear.

\section{Domains with collinear magnetization}\label{collinear magnetization}

\begin{figure*}[!t]
    \centering
    \includegraphics[width=0.9\textwidth]{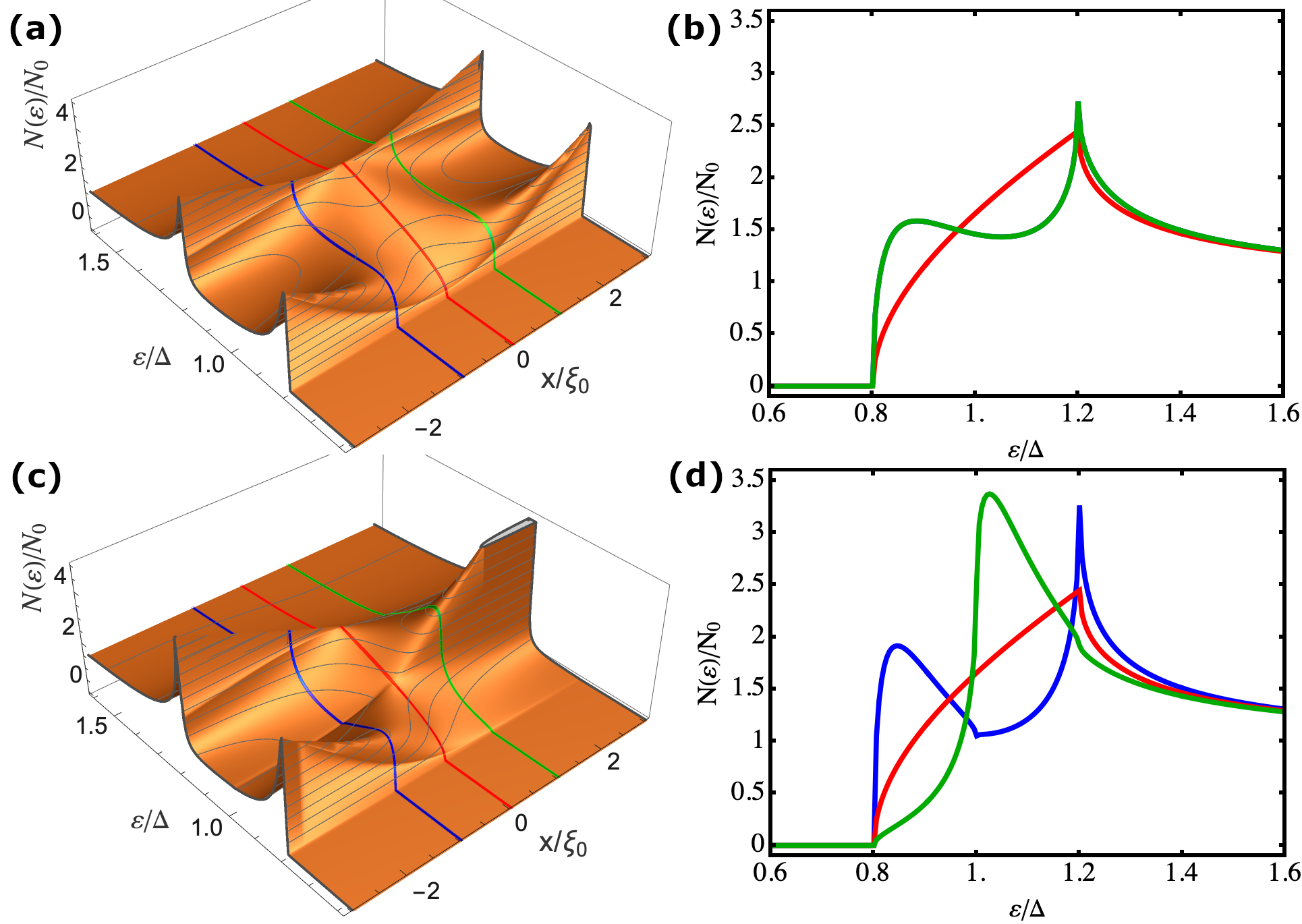}
    \caption{Local DoS of the S layer for (a,b) $h_l=0.2\Delta$ and $h_r=-0.2\Delta$ and (c,d) $h_l=0.2\Delta$ and $h_r=0$. The color lines in the right panels are taken at $x=-\xi_0$ (blue), $x=0$ (red) and $x=\xi_0$ (green).}\label{DOS plot 2}
\end{figure*}

If the magnetization of the two domains are  collinear, the problem can be greatly simplified.  
Firstly, only the component of the vector $\boldsymbol{V}$  parallel to the magnetization is non-zero. Without any loss of generality we assume that the magnetizations lie in the $z$ axis, such that $V_1=V_2=0$. In this case, Eq.~\eqref{Usadel equation 2} reads:
\begin{subequations}\label{Usadel-collinear}
\begin{align}
	D\theta''+2i\epsilon\sin{\theta}\cos{\theta_3}-2ih\cos{\theta}\sin{\theta_3}+2\Delta\cos{\theta}\cos{\theta_3} &= 0\\
	D\theta_3''+2i\epsilon\cos{\theta}\sin{\theta_3}-2ih\sin{\theta}\cos{\theta_3}-2\Delta\sin{\theta}\sin{\theta_3} &= 0\; ,
\end{align}
\end{subequations}
where $V_3=\sin{\theta_3}$. One can combine these equations to obtain two decoupled equations for each spin component:
\begin{equation}\label{Usadel-collinear-decoupled}
	D\theta_\pm''+2i\epsilon\sin{\theta_\pm} \mp 2ih\sin{\theta_\pm}+2\Delta\cos{\theta_\pm}=0\; ,
\end{equation}
where $\theta_\pm=\theta \pm \theta_3$ respectively describe the spin up and down components of the GF.

Since the problem  is  decoupled in spin space, one  can derive  equations in~\eqref{Usadel-collinear-decoupled} from  two independent Lagrangians: 
\begin{equation}\label{Lagrangian_pm}
	\mathcal{L}_\pm=\frac{D}{2}\left.\theta_\pm'\right.^2+2i\epsilon\cos{\theta_\pm} \mp 2ih\cos{\theta_\pm}-2\Delta\sin{\theta_\pm}\; .
\end{equation}
Because $\mathcal{L}_\pm$ do not depend explicitly on $x$, the following quantities are conserved in space:
\begin{equation}\label{E_pm}
	\mathcal{E}_\pm=\frac{D}{2}\left.\theta_\pm'\right.^2-2i\epsilon\cos{\theta_\pm} \pm 2ih\cos{\theta_\pm}+2\Delta\sin{\theta_\pm}\; .
\end{equation}
These expressions  can be evaluated at the bulk where the spatial derivative vanishes and the GF is given by the bulk solution [see Eq.~\eqref{bulk g,f}]. $\cos{\theta_\pm}$ and $\sin{\theta_\pm}$ are given by the spin components $\hat{g}$ and $\hat{f}$ respectively, where the $\pm$ sign corresponds to the up/down spin index
\begin{equation}\label{E_pm value}
	\mathcal{E}_\pm=2\sqrt{\Delta^2-(\epsilon\mp h)^2}=\frac{2\Delta}{\sin{\bar{\theta}_\pm}}\; .
\end{equation}
Here $\bar{\theta}_\pm$ are the value of $\theta_\pm$ at the bulk. In the following we omit the spin subscript to simplify the notation. Substituting equation~\eqref{E_pm} into~\eqref{E_pm value} and applying trigonometric identities, we arrive at
\begin{equation}\label{Usadel_int}
    \sin{\bar{\theta}}\frac{D}{8\Delta}\left.\theta'\right.^2=\sin^2{\frac{\theta-\bar{\theta}}{2}}\; .
\end{equation}
Equation~\eqref{Usadel_int} does not explicitly contain the independent variable $x$. Taking the square root on both sides of the equation we obtain a first order differential equation which can be integrated to obtain
\begin{equation}\label{tan}
    \tan{\frac{\theta-\bar{\theta}_{l/r}}{4}}=
    \begin{cases}
      c_l e^{x/\lambda_l}, & \ x \leq 0 \\
      c_r e^{-x/\lambda_r}, & x \geq 0
    \end{cases}\; ,
\end{equation}
where $\lambda^2_{\pm,l/r}=D/(2\sqrt{\Delta^2-(\epsilon \mp h_{l/r})^2})$ is chosen such that $\mathrm{Re}\{\lambda_{l/r}\}>0$, and  that the exponential functions decay away from the domain wall. $h_{l/r}$ is the value of the exchange field in the left ($x<0$) and right ($x>0$) domains.

From Eq.~\eqref{tan} one can obtain the spatial dependence of $\theta(x)$ by determining the constants $c_{l,r}$. 
For this we use the fact that  the GF and its derivative are  continuous at the domain wall. Applying this condition, we obtain the values of the constants in Eq.~\eqref{tan}
\begin{widetext}
\begin{equation}
    c_{l/r}=\mp\frac{\frac{\lambda_{l/r}}{\lambda_{r/l}}\left(1-\tan^2{\frac{\Delta\theta}{4}}\right)+1+\tan^2{\frac{\Delta\theta}{4}}-\sqrt{\left[\frac{\lambda_{l/r}}{\lambda_{r/l}}\left(1-\tan^2{\frac{\Delta\theta}{4}}\right)+1+\tan^2{\frac{\Delta\theta}{4}}\right]^2+\frac{4\lambda_{l/r}^2}{\lambda_{r/l}^2}\tan^2{\frac{\Delta\theta}{4}}}}{2\frac{\lambda_{l/r}}{\lambda_{r/l}}\tan{\frac{\Delta\theta}{4}}}\; ,
    \label{coeffs}
\end{equation}
\end{widetext}
where $\Delta\theta=\theta_r-\theta_l$,  and the upper and lower signs correspond to the left and right domains respectively. The sign of the square root on Eq.~\eqref{coeffs} is chosen such that the DoS is positive and the solution is physically meaningful. Setting the order parameter to zero in the right domain and the exchange fields to zero, we recover the results by Altland \textit{et al.}~\cite{Altland-2000} for a singlet S/N junction. Golubov \textit{et al.}~\cite{Golubov-2005} also followed a similar procedure to study the DoS at ferromagnetic and normal layers on S(FN) and S(FF) structures.

Equation~\eqref{tan} together with Eq.~\eqref{coeffs} determines the analytical solution for the two semi-infinite collinear domains. The local DoS is related to the GF through the expression
\begin{equation}
\begin{aligned}
	\frac{N(\epsilon)}{N_0} &=\frac{1}{2}\mathrm{Re}\{\mathrm{Tr}\; \hat{g}(\epsilon)\}\\ &=\frac{1}{2}\mathrm{Re}\{\cos{\theta_+}+\cos{\theta_-}\}\; .
\end{aligned}
\end{equation}
As first example, we assume that the magnetizations of the two domains are opposite in direction but equal in amplitude ($h_l=-h_r=h$). In this case the DoS at the domain wall ($x=0$) has a simple form 
\begin{equation}
	\frac{N(\epsilon)}{N_0}=\mathrm{Re}\left\{\frac{\sqrt{\Delta^2-(\epsilon-h)^2}-\sqrt{\Delta^2-(\epsilon+h)^2}}{2ih}\right\}\; , 
\end{equation}
which leads to the red curves in Fig.~\ref{DOS plot}.
This analytical result coincides with the numerical result obtained in Refs.~\cite{Strambini-2017} and~\cite{Aikebaier-2019} for a narrow domain wall between two collinear domains. In Fig.~\ref{DOS plot} we show the spatial dependence of the  DoS for spin-up electrons  in the antiparallel magnetization configuration. Far form the domain wall the coherent peaks of the DoS are well defined (dashed lines). The presence of the domain wall smears the peak. In Fig.~\ref{DOS plot 2}(a) we show the spatial dependence of the full DoS. Specifically,  Fig.~\ref{DOS plot 2} (b)  shows the DoS at the values of $x$ indicated by the colored lines in panel (a). The magnitude of the exchange field is the same on both domains so the total DoS is symmetric with respect to $x=0$. For large enough distances  away from the domain wall  the BCS peak is shifted by the exchange field to $\epsilon=\Delta \pm h$. Near the domain wall, there is a crossover between the position of the spin up/down peaks over a length scale of the order of the superconducting coherence length $\xi_0=\sqrt{D/\Delta}$. Notice that  around the domain wall, the inner peak is broader and lower than the outer peak [see Fig.~\ref{DOS plot 2}(b)], but the gap edge remains, as expected, at $\epsilon<\Delta - h$. 

A second interesting example is  when the exchange field is only finite in one of the regions ($x<0$). This  corresponds to a S layer only partly covered by the the FI layer. In Fig.~\ref{DOS plot 2}(c) we show the local DoS in this case.  The spin-split DoS at $x\ll -\xi_0$ evolves  into the usual BCS DoS at $x\gg \xi_0$, over the length $\xi_0$ around the domain wall.  The splitting of the DoS peaks does not decrease smoothly as one would expect in a system where the exchange field is suppressed gradually over a length much larger than $\xi_0$. Namely, the inner peak is smeared in a similar way to the anti-parallel magnetization case [Fig.~\ref{DOS plot 2}(b)], such that the DoS has the same ``shark-fin" shape right at  $x=0$ [red curve on Fig.~\ref{DOS plot 2}(d)]. All  above predictions could be proven by performing local tunneling spectroscopy measurements.

\section{Non-collinear magnetization}\label{arbitrary magnetization}
In the previous section we focused on the collinear magnetization case in which it was possible to decouple the components of the Usadel equation. In that case, we can find conserved quantities, Eq. (\ref{E_pm value}), and obtain analytically expressions for the GF. If the magnetizations are  non-collinear the system lacks enough symmetries to reduce the number of coupled equations. Nonetheless, it is possible to solve the Usadel equation~\eqref{Usadel equation 2} analytically in the weak superconducting or weak exchange field limits, as discussed in the next subsections. Later below, we study the two domain situation for an arbitrarily large exchange field numerically~\footnote{Domain walls with non-collinear magnetization may induce  stray fields that could affect superconductivity. In this work, we neglect the orbital effects of stray fields and assume that the superconducting gap is homogeneous in the S layer. The stray fields depend on the geometry and size of the magnetic layer, so we assume that the FI geometry is such that it minimizes the magnetostatic energy and, therefore, the stray fields. For any angle $\alpha$, the stray field can be minimized if the orientation of the domains is chosen to be $\pm\alpha/2$ with respect to the $x$ axis.}.

\subsection{Weak superconductivity}\label{weak superconductivity}
If the superconductor is close to the critical temperature $T_c$, the Usadel equation~\eqref{Usadel equation 2} can be linearized for small order parameter ($\Delta \ll T,h$). This limit is very illustrative to understand the length scales involved in the system.

Near $T_c$ the GF can be approximated by $\check{g}=\tau_3+\hat{f}\tau_1$. The linearized Usadel equation determines the anomalous GF $\hat{f}$
\begin{subequations}\label{Usadel linearized Delta}
\begin{align}
    \frac{D}{2}f_0''+i\epsilon f_0-i\boldsymbol{h}\cdot\boldsymbol{f}+\Delta &= 0\\
    \frac{D}{2}\boldsymbol{f}''+i\epsilon \boldsymbol{f}-i\boldsymbol{h}f_0 &= 0\; ,
\end{align}
\end{subequations}
where the spin structure is
\begin{equation}\label{f_structure}
    \hat{f}=f_0+\sum_{j=1,3} f_j\sigma_j\; ,
\end{equation} 
where $f_0$ is the singlet and $f_j$, $j=1,2,3$ the triplet components. 
For the two semi-infinite domain structure considered in this work, the solution to Eq.~\eqref{Usadel linearized Delta} is given by
\begin{subequations}\label{f solution}
\begin{align}
    f_0 =& \overline{f}_0+c_+e^{-q_+ |x|}+c_-e^{-q_- |x|}\\
    \boldsymbol{f} =& \overline{\boldsymbol{f}}+i\boldsymbol{h}\sum_{j=\pm} \frac{c_j}{q_j^2 D/2+i\epsilon}e^{-q_j |x|}+\boldsymbol{d}e^{-q |x|}\; ,\label{eq:lin_tripletpara}
\end{align}
\end{subequations}
where $\overline{f}_0$ and $\overline{\boldsymbol{f}}$ are the asymptotic  values at $x=\pm \infty$ of the singlet and triplet components 
respectively, and $q_{\pm}^2=-2i(\epsilon \mp h)/D$, $q^2=-2i\epsilon/D$.  The triplet can be written as the sum of the component parallel to the local exchange field, second term in Eq. (\ref{eq:lin_tripletpara}), and the component orthogonal to it proportional to the vector $\boldsymbol{d}$, with  $\boldsymbol{h}\cdot\boldsymbol{d}=0$. The component perpendicular to the local exchange field decays away from the domain wall over the length $\xi_\epsilon= \mathrm{Re}\{q\}^{-1}$, whereas the correction to the bulk (parallel)  solution is significant at distances less than  $\xi_h= \mathrm{Re}\{q_+\}^{-1}=\mathrm{Re}\{q_-\}^{-1}$.

\subsection{Weak exchange field}\label{linearized magnetization}
Another analytical limiting case is the case of a weak  exchange field  ($|\boldsymbol{h}| \ll \Delta$). In this case one can linearize the Usadel equation~\eqref{Usadel equation 2} and solve the system for an arbitrary magnetization texture. In zeroth order in $\boldsymbol{h}$, only the singlet component of the GF is finite and it is given by  Eqs.~(\ref{bulktheta}-\ref{bulk g,f}) setting   $\boldsymbol{h}=0$
\begin{subequations}
\begin{align}
    \tan{\theta}=\frac{\Delta}{-i\epsilon}\\
    V_0=1\; .
\end{align}
\end{subequations}
To first order in $\boldsymbol{h}$, both $\theta$ and $V_0$ are not corrected.  Whereas the triplet vector  $\boldsymbol{V}$  is determined by
\begin{equation}\label{Usadel equation linearized}
	\boldsymbol{V}''-\lambda^{-2} \boldsymbol{V} = \frac{2i\Delta}{D\sqrt{\Delta^2-\epsilon^2}}\boldsymbol{h}(x')\; ,
\end{equation}
where $\lambda^2=D/(2\sqrt{\Delta^2-\epsilon^2})$ ($\mathrm{Re}\{\lambda\}>0$) is the energy dependent coherence length.  The solution of this equation can be written as 
\begin{equation}
    \boldsymbol{V}=\int dx' G(x,x')\frac{2i\Delta}{D\sqrt{\Delta^2-\epsilon^2}}\boldsymbol{h}
\end{equation}
where $G(x,x')$ is the Green's function  of the differential equation~\eqref{Usadel equation linearized} determined by
\begin{equation}\label{GF equation}
    \left(\partial_x^2-\lambda^{-2}\right) G(x,x')=\delta(x-x')\; .
\end{equation}
%Fourier transforming Eq.~\eqref{GF equation} in the variable $x-x'$,
%\begin{equation}
%    G(k)=\frac{-1}{k^2+\lambda^{-2}}\; .
%\end{equation}
%Applying the inverse Fourier transform, we arrive at
Solving Eq.~\eqref{GF equation}, we arrive at
\begin{equation}\label{V_1}
    \boldsymbol{V}(x)=\frac{-i\Delta}{\sqrt{2D}(\Delta^2-\epsilon^2)^{3/4}}\int dx' e^{-|x-x'|/\lambda}\boldsymbol{h}(x')\; .
\end{equation}
This result shows  explicitly the spatial dependence of the  triplet vector. It  is determined by the exchange field averaged over the length $\lambda$.  
For example if the spatial variation of the exchange field $\boldsymbol{h}(x)$ is slower  than the  length $\lambda$ then  the  vector $\boldsymbol{V}$ is locally parallel to the exchange field. 
In particular, in the case of 
two magnetic domains separated by a smooth (respect to  the length $\lambda$) domain wall,
the vector $\boldsymbol{V}$  is always aligned with the local field $\boldsymbol{h}$.

In this work we are mainly interested in sharp domain walls , {\it i.e.} domain-walls with sizes much smaller than $\lambda$. If we model such  situation by a step-like exchange field with $\boldsymbol{h}=\boldsymbol{h}_l\theta(-x)+\boldsymbol{h}_r\theta(x)$, then the triplet vector at the left and right side of the domain wall can be  obtained from Eq. (\ref{V_1}):
\begin{equation}
    \boldsymbol{V}_{l/r}=\frac{-i\Delta}{2(\Delta^2-\epsilon^2)}\left[2\boldsymbol{h}_{l/r}+(\boldsymbol{h}_{r/l}-\boldsymbol{h}_{l/r})e^{-|x|/\lambda}\right]\; .
\end{equation}
As expected, at distances much  larger than $\lambda$ from the domain wall $\boldsymbol{V}$ is parallel to the local exchange field. In contrast, the transverse component to the field is  maximized at the domain wall and decays over $\lambda$ away from it. 
The above analytical results are obtained for weak exchange fields. In the next section we consider arbitrarily strong exchange field.

\subsection{Arbitrary exchange field}\label{arbitrary direction}

\begin{figure}[!t]
    \centering
    \includegraphics[width=0.99\columnwidth]{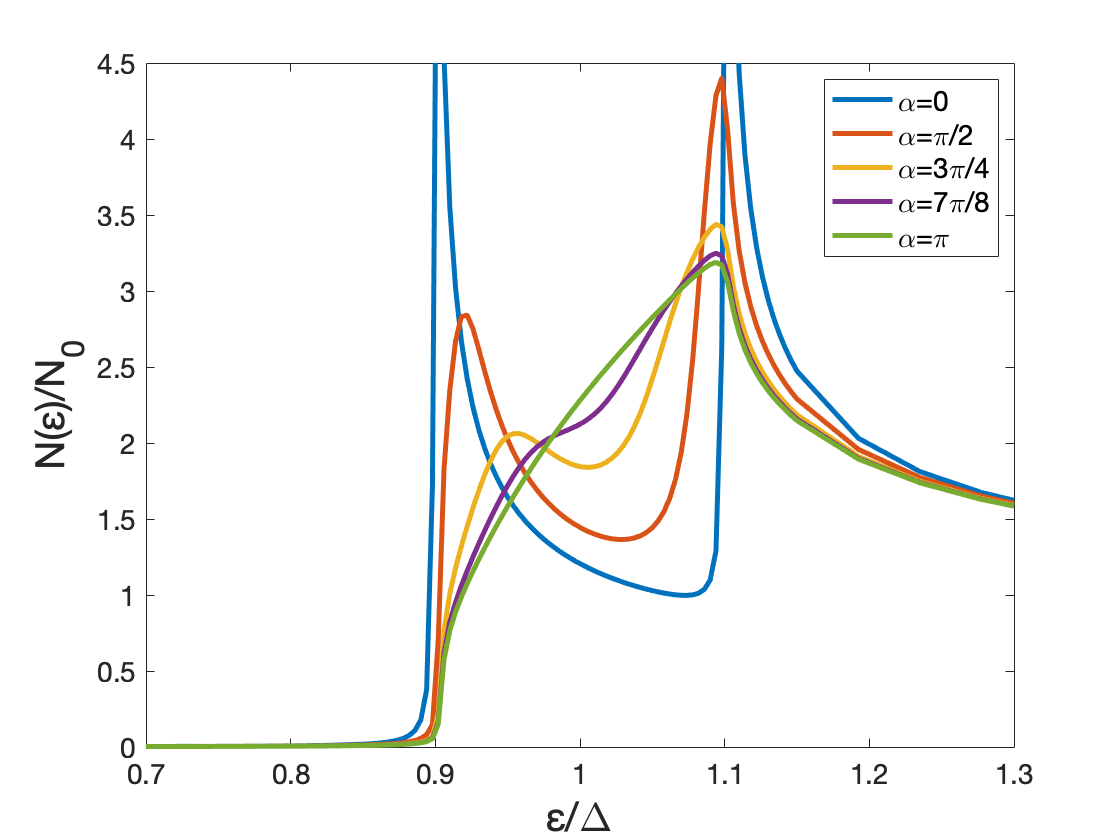}
    \caption{Density of states at the domain wall for different values of the angle $\alpha$ between the domains' magnetizations.}\label{DOS_plot_x_0}
\end{figure}

\begin{figure*}[!t]
    \centering
    \includegraphics[width=\textwidth]{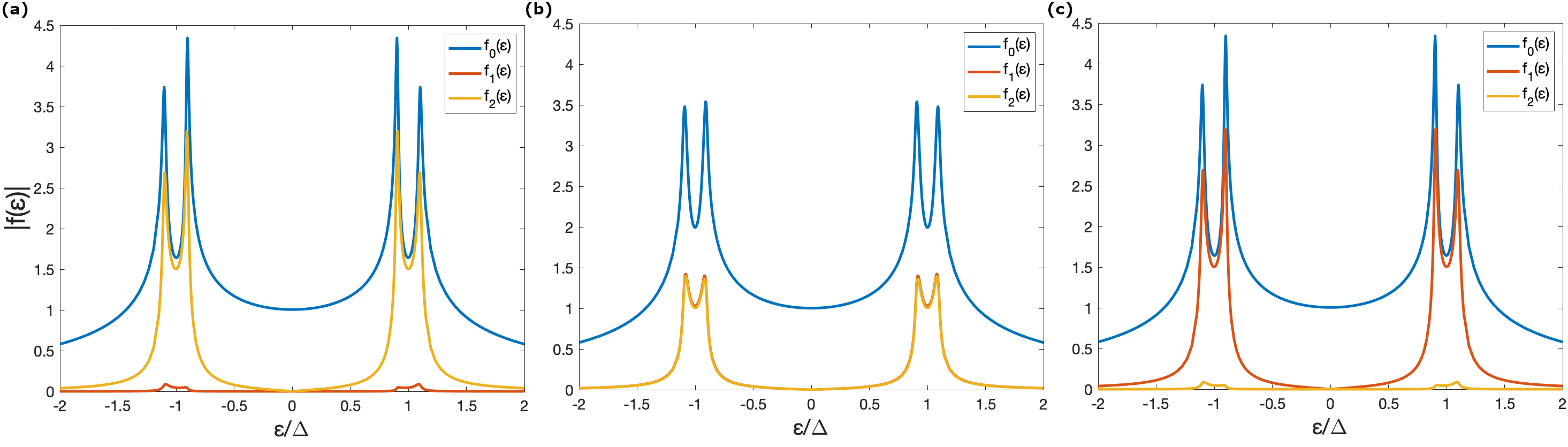}
    \caption{The spectral weight of  the singlet and triplet components of $\hat{f}$ at different points in the superconductor:  (a) $x=-5\xi_0$, (b) $x=0$ and (c) $x=5\xi_0$. We have chosen $\alpha=\pi/2$,  $L=10\xi_0$ and $h_{l/r}=0.1\Delta$.}\label{f_plot}
\end{figure*}

In this section we consider two domains with an arbitrarily large exchange field and arbitrary angle between the domains magnetization and solve numerically  the Usadel equation. For this it is convenient to differentiate twice Eq.~\eqref{normalization} and substitute the result  into Eq.~\eqref{Usadel equation 2}.  We thus arrive at ~\cite{Baker-2016,Aikebaier-2019}
\begin{subequations}\label{Usadel equation 3}
\begin{align}
	& D\theta''+2i\epsilon\sin{\theta}V_0-2i\cos{\theta}\boldsymbol{h}\cdot\boldsymbol{V}+2\Delta\cos{\theta}V_0=0\\
	\begin{split}
	    &D \boldsymbol{V}''+ D \boldsymbol{V}(\left.V_0'\right.^2+\left.\boldsymbol{V}'\right.^2) +2i\sin{\theta}((\boldsymbol{h}\cdot\boldsymbol{V})\boldsymbol{V}-\boldsymbol{h})\\
	    & \hspace{90pt}
	    -2(-i\epsilon\cos{\theta}+\Delta\sin{\theta})V_0\boldsymbol{V}=0
	\end{split}\\
	\begin{split}
	    &D V''_0+ D V_0(\left.V_0'\right.^2+\left.\boldsymbol{V}'\right.^2) +2i\sin{\theta}\boldsymbol{h}\cdot\boldsymbol{V}V_0\\
	    & \hspace{75pt}
	    +2(-i\epsilon\cos{\theta}+\Delta\sin{\theta})(1-V_0^2)=0\; .
	\end{split}
\end{align}
\end{subequations}
We solve the above equations numerically for a S layer of finite length $L$. The domain wall is located at $x=0$.  The spectral current vanishes at the boundaries  with vacuum.  In the generalized $\theta$-parametrization this translates into the following boundary conditions for Eq.~(\ref{Usadel equation 3})
\begin{subequations}\label{boundary conditions 3}
\begin{align}
    \left.\theta'\right|_{x=\pm L/2}&=0\\
    \left.\boldsymbol{V}'\right|_{x=\pm L/2}&=0\\
    \left.V'_0\right|_{x=\pm L/2}&=0\; .
\end{align}
\end{subequations}

In Fig.~\ref{DOS_plot_x_0} we show the computed total DoS at the domain wall for domains with the same exchange field  magnitude $h=0.1 \Delta$ and different  orientations [see Fig.~\ref{S-FI schematic}]. In the $\alpha=0$ case, the exchange field is uniform along the sample, so the DoS is the  homogeneous  spin-split BCS. Both peaks are broadened and lowered by increasing $\alpha$, and the DoS exhibits the ``shark-fin"  when the magnetizations are anti-parallel. The spin-splitting is still visible  up to values of $\alpha \approx 7\pi/8$. %The depth of the minimum between the two coherence peaks is a good indicator of the relative orientation of the domains.  It is very deep in the homogeneous case while the depth decreases monotonically with increasing $\alpha$. 

\section{Triplet pair correlations in FI/S structures and their detection}\label{entangled triplet pairs}

In the previous sections, we have analyzed the quasiparticle spectrum. Here we focus on  another aspect of the FI/S structures: 
the superconducting triplet pair-correlations. These appear due to the finite interfacial exchange field that converts conventional singlet into triplet pairs~\cite{bergeret2001long,Bergeret:2005}.

Within our model, pair correlations are described by the anomalous component $\hat f$ introduced in Eq.~\eqref{eq:gen_str_g}, which in the spin-space has the general structure given by Eq.~\eqref{f_structure}. Because we consider the strict diffusive limit, all components of $\hat f$ are isotropic in momentum (s-wave symmetry). From the Fermi statistics for Fermion pairs it follows that $f_0$ is an even function of frequency whereas $f_j$ are odd~\cite{berezinski1974new,Bergeret:2005,Tanaka-Odd-Frequency,linder2019odd}.  
The following association  between the different components of the condensate and the spin state of  electron pairs can be made~\cite{eschrig2011spin}
\begin{align}
    & (\uparrow\downarrow-\downarrow\uparrow)\leftrightarrow 2 f_0 \label{f0}\\
    -&(\uparrow\uparrow - \downarrow\downarrow)\leftrightarrow 2 f_1\\
    & (\uparrow\uparrow + \downarrow\downarrow) \leftrightarrow 2i f_2\\
    & (\uparrow\downarrow+ \downarrow\uparrow)\leftrightarrow 2 f_3 \, . \label{f3}
\end{align}
In other words, each triplet component of the condensate is associated with maximally entangled states. In a conventional BCS superconductor, only the singlet component $f_0$ is finite. Triplet components are finite in the presence of an exchange field. In a homogeneous case, we choose the spin quantization axis along the magnetization direction, e.g., the $z$ axis. The only finite components of the condensates are, in this case, $f_0$ and $f_3$. All triplet components may appear in a multidomain situation with arbitrary magnetization directions.

\begin{figure*}[!t]
    \centering
    \includegraphics[width=\textwidth]{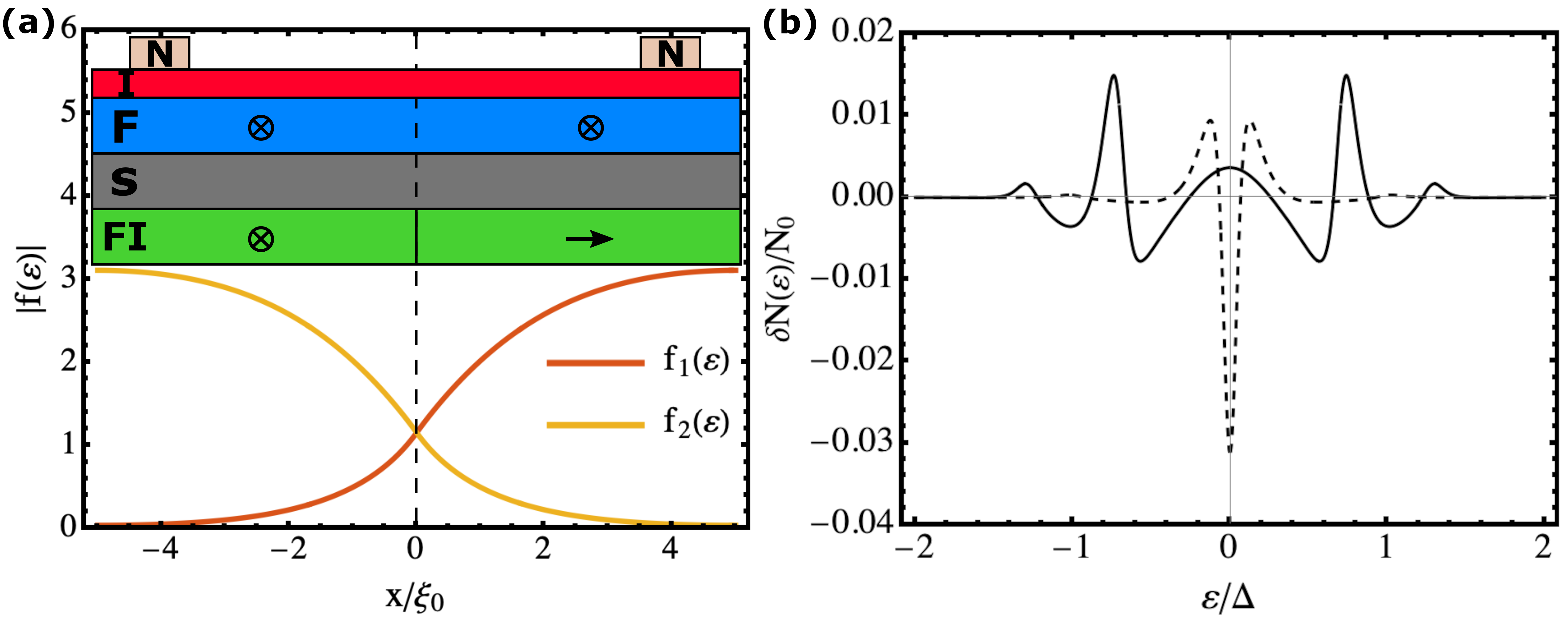}
    \caption{(a) Spatial dependence of triplet correlations for energy $\epsilon=\Delta-h$. We show the proposed geometry to detect the triplet correlations in the inset. A F layer is placed on top of a  S layer; if the F layer is thick enough, only the triplet correlations perpendicular to the magnetization of the F layer will propagate along the ferromagnet. The long-range triplet correlations manifest as a zero-energy peak on the local DoS, measured through tunnel differential conductance measurements with a normal metal probe  (N).
    (b) The correction to the DoS of the ferromagnet at the F/I interface, see panel (a), far to the right of the domain wall (solid line). The deviation from the normal DoS is due to the penetration of the long-range component of the triplet condensate. For comparison, we show the  DoS of an N layer in contact with a conventional singlet superconductor (dashed line). At zero energy,  the singlet (triplet) component induced in the N (F) layer is real (purely imaginary), resulting in a negative (positive) correction to the DoS. The parameters used in the plot are $h=0.3\Delta$, $\gamma=5\xi_0$ and $t=3\xi_0$. The DoS of the F case is normalized by  $(\Delta/\Gamma)^2$ for comparison.}\label{DOS_F_plot}
\end{figure*}

Here we study the singlet and triplet correlations in a FI/S bilayer with two non-collinear domains, see Fig. \ref{S-FI schematic}.
We assume that $\alpha=\pi/2$ such that two components $f_1$, $f_2$ are finite. The length of S  is $L=10\xi_0$. In Fig.~\ref{f_plot} we show the spatial dependence  of the singlet and triplet components of $\hat{f}$ for all energies, calculated numerically.
All condensate components show peaks at  $|\epsilon|=\Delta \pm h$ and decay to zero at energies much larger than the gap. Inside the gap 
the amplitude of the singlet is of the order of 1. In contrast, at $\epsilon=0$ the triplet components are of the order of $\Gamma/\Delta$, where $\Gamma$ is the Dynes parameter~\cite{Dynes:1978} describing inelastic scattering. They increase linearly at small energies and become comparable to the singlet component within the range  $|\epsilon| \in [\Delta-h,\Delta+h]$. Far away from the domain wall, only the triplet component parallel to the local exchange field is finite.   Both components, $f_1$ and $f_2$, have the same magnitude at the domain wall, as anticipated from our analytical result, Eq. (\ref{V_1}). In Fig.~\ref{DOS_F_plot}(a) we show the spatial dependence of the triplet correlations at $\epsilon=\Delta-h$. The length over which the triplet components change is of the order of the coherence length. 

A natural question is how to detect the triplet components in this type of system. This can be achieved for example through spin-polarized spectroscopy~\cite{Bobkova_2019}. Another way to detect the triplet components is to place a ferromagnetic layer (F)  on top of a  superconductor.   The DoS of the F layer is modified by the superconducting correlations induced via the proximity effect. Such modification can be measured by a normal tunneling probe. In the case of a weak proximity effect  we can linearize the Usadel equation in the F region. The DoS in the ferromagnet is then given by (see Appendix for details): 
\begin{equation}\label{DOS linearized}
    \frac{N(\epsilon,x,z)}{N_0}=1-\frac{1}{4}\mathrm{Re}\{ \mathrm{Tr} \; \hat{f}^2(\epsilon,x,z)\}\; .
\end{equation}
The second term is the correction to the DoS due to the proximity effect. Because of the trace over spin, this term has two contributions:  One proportional to the square of the singlet component and one to the sum of the squares of the triplet components. 
The singlet component is real at low energies, so its correction to the DoS is negative. This explains that if S is a singlet superconductor and F is a normal layer (no exchange and hence no triplet), the  DoS is suppressed at $\epsilon=0$, see dashed line in  Fig.~\ref{DOS_F_plot}(b). On the other hand, in the presence of an exchange field, the triplet component at $\epsilon=0$ is purely imaginary [see Eq.~\eqref{GF F}] and hence its contribution to the DoS, according to Eq.~\eqref{DOS linearized}, is positive. Thus, the sign of the correction of the DoS at $\epsilon=0$, is determined by competition between singlet and triplet amplitudes~\cite{Yokoyama:2007}.

%First, we assume that the thickness of the attached layer is small compared to the superconducting coherence length and that the exchange field is negligible (normal metal limit). The superconducting correlations induce a minigap of width $\tilde{\Delta} \approx \epsilon_b=D/(2\gamma t)$, where $t$ is the thickness of the attached layer, $\gamma$ is a parameter describing the interface resistance and $\epsilon_b$ is an energy scale related to the interface. The DoS for this configuration is shown in Fig.~\ref{DOS_F_plot}(b). However, if the exchange field is grater than $\tilde{\Delta}$, the minigap is closed leading to a zero-energy peak of the local DoS~\cite{Yokoyama:2007}. These zero-energy corrections to the DoS are attributed to the short-range singlet and triplet corrections. If the superconductor-ferromagnet interface is very resistive, the proximity effect is very weak and the DoS of the metal is given by Eq.~\eqref{DOS linearized}. At low energies, the singlet component of the GF induced in the normal metal is real, so the DoS has a negative correction. In the F layer, however, the odd-frequency spin-triplet pairing states are dominant. According to Eq.~\eqref{GF F}, at $\epsilon=0$ the triplet component of the GF induced in the F layer is imaginary, resulting in a positive correction of the DoS.

\begin{figure*}[!t]
    \centering
    \includegraphics[width=0.8\textwidth]{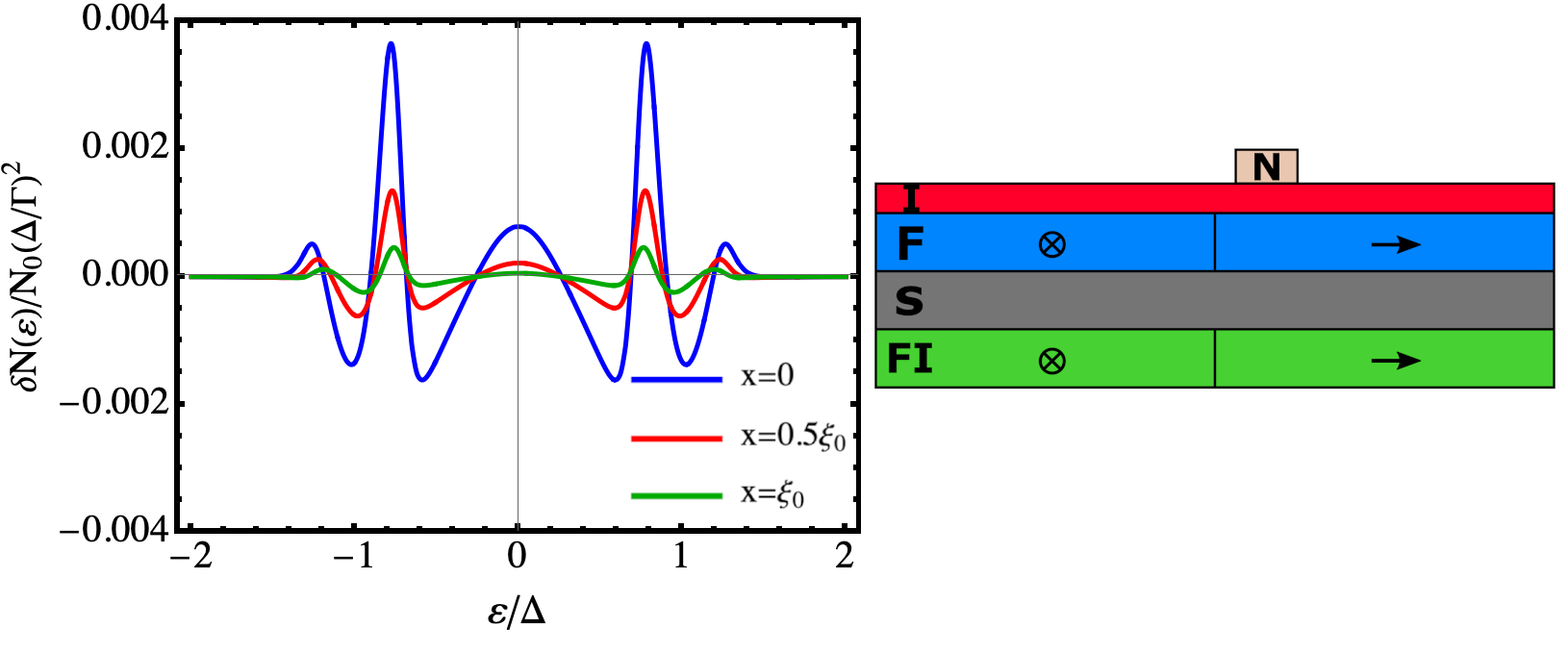}
    \caption{Correction to the DoS of the ferromagnet at the F/I interface at different distances from the domain wall. In this setup, the ferromagnet magnetization is aligned with respect to  the magnetization of the FI.  The parameters used are $h=0.3\Delta$, $\gamma=5\xi_0$ and $t=3\xi_0$.}\label{DOS_F_plot_2}
\end{figure*}

In order to separate the triplet from the singlet component, we propose a setup as the one sketched in  Fig.~\ref{DOS_F_plot}. Due to the presence of the FI, triplet pairs are induced in the superconductor, as described above. In order to filter out the singlet correlations, an F layer with a magnetization non-collinear to the FI is placed, see inset of Fig.~\ref{DOS_F_plot}(a). The singlet component and triplet parallel to the F magnetization  (short-range triplet) decay over the magnetic length $\sim \kappa_F^{-1}$. In contrast,  the triplet component orthogonal to the magnetization of F  (long-range triplet), 
decays over the length $\sim \kappa_\epsilon^{-1}$ [see Appendix~\ref{tunneling differential conductance}]. 
 Thus, by choosing the thickness $t$ of the F layer such that $\kappa_F^{-1} \ll t \ll \kappa_\epsilon^{-1}$, the DoS of F at the tunneling barrier will be only corrected by the long-range triplet component. This situation can be realized by using  F layers with a strong exchange field, such as Co or Fe.

In Appendix~\ref{tunneling differential conductance} we compute the correction to the density of states in the ferromagnet. In the two domain situation studied above, when the F layer is placed above the right domain far from the domain wall, see Fig.~\ref{DOS_F_plot}(a), the triplet component in F  at zero-energy  is purely imaginary, Eq.~\eqref{GF F}, so according to Eq.~\eqref{DOS} there is a positive correction to the DoS
\begin{equation}\label{DoS F}
    \frac{N(0,\infty,t)}{N_0}=1+\frac{\epsilon_b^2 h^2 \Delta^2}{2 (\Delta^2-h^2)^3}\; ,
\end{equation}
where  $\epsilon_b=D/(2\gamma t)$ is an energy scale related to the interface transparency, and  $\gamma$ is a parameter describing the interface resistance. The solid line in  Fig.~\ref{DOS_F_plot}(b) shows the DoS of the F layer at the tunneling barrier computed for all energies. One sees a local maximum at $\epsilon=0, $ and also maxima at $|\epsilon|=\Delta \pm h$  related to the triplet peaks shown in Fig.~\ref{f_plot}. In this way, the existence of triplets generated in the spin-split superconductor  can be   demonstrated by performing 
tunneling spectroscopy, with the  normal electrode probe, see Fig~\ref{DOS_F_plot}(a).

Finally, we consider a F layer consisting of two domains that are collinear to the adjacent FI domains [see Fig.~\ref{DOS_F_plot_2}]. 
This situation may correspond to the case that the magnetic coupling of the  F and FI  leads to local collinear magnetizations. According to our previous analysis, triplet correlations of both kinds are present in the S near the domain wall. In other words,  long-range triplet correlations will be present in certain positions of the F/I interface and affect the local DoS. In Fig.~\ref{DOS_F_plot_2} we show the correction to the DoS at different points of the F/I interface. The zero-energy peak appears at regions close to the domain wall. The peak vanishes when moving away from the domain wall. Such measurements could be done with the help of the STM technique and may reveal the magnetic texture of the system. Another possible setup to isolate the odd-frequency correlations at zero energy are S/N bilayers with a spin-active interface~\cite{Linder-2009,Linder-2010}.

\section{Conclusion}\label{conclusion}

In this work, we have studied the spectral properties of superconductor-ferromagnetic insulator bilayers in the presence of a domain wall separating two magnetic domains. 
In the first part, we focus on the quasiparticle spectrum and analyze how the density of states of the superconductor is affected by the magnetic configuration. In the case of two semi-infinite domains with collinear magnetization and a sharp domain wall between them,   it is possible to find two integrals of motion that allow for an analytical solution of the Usadel equation. With the help of this solution, we determine the local DoS of the superconductor for different magnitudes of the exchange field. At the domain wall, the DoS exhibits a ``shark-fin" shape. This feature appears when the domain magnetizations are antiparallel or when one of the domains has a negligible small exchange field. 
We have also studied FI layers with non-collinear magnetization direction. We show that near the domain wall, the spin-splitting is quite robust with respect to the relative angle $\alpha$ between the magnetizations, but the heights of the coherent peaks are significantly affected by it. All these predictions can be verified by local tunnel spectroscopy experiments, which will reveal information about the local magnetic configuration of the FI.

In the second part, we have analyzed the spectral properties of the singlet and triplet components of the superconducting condensate in the S layer. We have found an analytical expression for the quasiclassical Green's function in the presence of an arbitrary magnetic texture in the FI in the case of a weak exchange field. Our expression reveals how the local exchange field spatially determines the triplet components induced in the superconductor. For arbitrary strength of the exchange interaction, we have determined the singlet and triplet components numerically in the presence of a sharp domain wall. 
We propose different ways of detecting the triplet correlations using a FI/S/F junction, where F is a ferromagnetic metal and a tunneling probe at the outer F interface.  The presence of the triplet component manifests itself as a zero bias maximum in the tunneling differential conductance. The proposed setup can then be used as a source of spin-triplet pairs, whose entanglement can be proved in experiments using quantum dots as pair splitters~\cite{Hofstetter-2009}.

\section*{Acknowledgements}
We thank S. Ili\'{c} for useful discussions. 
This work was partially  funded by the Spanish Ministerio de Ciencia, Innovación y Universidades (MICINN) through Project PID2020-114252GB-I00 (SPIRIT), and EU’s Horizon 2020 research and innovation program under Grant Agreement No. 800923 (SUPERTED).
A.H. acknowledges funding by the University of the Basque Country (Project PIF20/05).

\vspace{1cm}

\begin{appendices}
\numberwithin{equation}{section}

\section{Correction to the tunneling differential conductance in the ferromagnet}\label{tunneling differential conductance}
In this appendix we show how the tunneling differential conductance measured on top of the F layer [see Fig.~\ref{DOS_F_plot}(a)] is affected by the by the leakage of the superconducting condensate into the ferromagnet.

The GF on a diffusive ferromagnet satisfies the Usadel equation~\eqref{usadel_equation} with $\Delta=0$. If the transmission coefficient of the S/F interface is very low, the proximity effect in the F layer is weak and the Usadel equation can be linearized as
\begin{subequations}\label{Usadel F}
\begin{align}
	\partial^2_{zz} f_0+i\kappa_\epsilon^2 f_0 -i\kappa^2_F f_2 &= 0\\
	\label{Usadel F triplet}
	\partial^2_{zz} \boldsymbol{f}+i\kappa_\epsilon^2 \boldsymbol{f} -i\kappa^2_F f_0 \hat{\boldsymbol{y}} &= 0\; ,
\end{align}
\end{subequations}
where $\kappa_\epsilon^2=2\epsilon/D$, $\kappa_F^2=2h_F/D$ and $h_F$ is the field of the ferromagnet. Here we have assumed that the magnetization direction of the F layer lies on the $y$ axis.

The S/F interface is described by the linearized Kuprianov-Lukichev condition~\cite{Kuprianov-Lukichev}
\begin{subequations}\label{Kuprianov-Lukichev}
\begin{align}
	\gamma\left.\partial_z f_0\right|_{z=0} &= -f_{S,0}\\
	\gamma\left.\partial_z \boldsymbol{f}\right|_{z=0} &= -\boldsymbol{f}_S\; .
\end{align}
\end{subequations}
Here, $\gamma=\sigma_F R_b$ is the parameter describing the barrier strength, where $R_b$ is the normal-state tunneling resistance per unit area and $\sigma_F$ is the conductivity of the ferromagnet. The anomalous GF on the S layer is given by $\hat{f}_S=f_{S,0}+\boldsymbol{f}_S\cdot\boldsymbol{\sigma}$.

We assume that the thickness $t$ of the F layer is much longer than the coherence length in the ferromagnetic layer $\kappa_F t \gg 1$. In the long-junction regime the condensate function is mediated primarily by the long-range triplet superconducting correlations~\cite{Bergeret:2005,Buzdin:2011,Samokhvalov:2019}, whereas the singlet and short-range triplet correlations decay over the length $\kappa_F^{-1}$. At the outer interface of the F layer, the condensate function is given by the only long-range component $f_1$. Solving the Usadel equation~\eqref{Usadel F triplet} we obtain
\begin{equation}\label{GF F}
    f_1(\epsilon,x,t)=\frac{f_{S,1}(\epsilon,x)}{\frac{1-i}{\sqrt{2}}\kappa_\epsilon \gamma \sinh{\left( \frac{1-i}{\sqrt{2}} \kappa_\epsilon t \right)}}\; .
\end{equation}

In the case of weak proximity effect, the DoS of the ferromagnet is given by Eq.~\eqref{DOS linearized}. Using Eq.~\eqref{GF F}, we arrive at
\begin{equation}\label{DOS}
    \frac{N(\epsilon,x,t)}{N_0}=1-\frac{1}{2}\mathrm{Re}\left\{ \frac{i f_{S,1}(\epsilon,x)^2}{\gamma^2 \kappa_\epsilon^2 \sinh^2{\left( \frac{1-i}{\sqrt{2}} \kappa_\epsilon t \right)}} \right\}\; ,
\end{equation}
where the anomalous GF of the superconductor $f_{S,1}(\epsilon,x)$ is obtained by solving Eqs.~(\ref{Usadel equation 3}-\ref{boundary conditions 3}). If the S layer is a homogeneous superconductor with a exchange field along the $x$ direction, the DoS is given by Eq.~\eqref{DoS F}.

\end{appendices}

%\nocite{*}
%\bibliography{biblio}\label{LastBibItem}


\end{document}